\documentclass{elsart}
\usepackage{epsfig}

\newcommand{\asc}[1]%
             {$\langle${#1}$\rangle$}

\begin{document}

\begin{frontmatter}
\title{Impact of Varying Atmospheric Profiles on Extensive Air Shower Observation: \\
 - Atmospheric Density and Primary Mass Reconstruction - }

\author[Uni]{B.
Keilhauer}\renewcommand{\thefootnote}{\fnsymbol{footnote}}\footnote{Corresponding
author. \em{E-mail-Address:} bianca.keilhauer@ik.fzk.de}, 
\author[Uni,FZK]{J. Bl\"umer},
\author[FZK]{R. Engel},
\author[FZK]{H.O. Klages},
\author[FZK,Polen]{M. Risse}

\address[Uni]{Universit\"at Karlsruhe, Institut f\"ur Experimentelle Kernphysik, Postfach
3640, 76021 Karlsruhe, Germany}
\address[FZK]{Forschungszentrum Karlsruhe, Institut f\"ur Kernphysik, Postfach 3640, 76021
Karlsruhe, Germany}
\address[Polen]{H.~Niewodnicza\'nski Institute of Nuclear Physics, Polish Academy of
Sciences, ul.~Radzikowskiego 152, 31-342 Krak\'ow, Poland}

\begin{abstract}

The longitudinal profile of extensive air showers is sensitive to the energy and type / mass 
of the primary particle. One of its characteristics, the atmospheric depth of 
shower maximum, is often used to reconstruct the elemental composition of primary cosmic rays.
In this article, the impact of the atmospheric density profile on the reconstruction of 
the depth of maximum, as observed in fluorescence light measurements, is investigated. 
We consider in detail the atmospheric density profile and its time 
variations at the site of the southern Pierre Auger Observatory, using data that were 
obtained from meteorological radio soundings.
Similar atmospheric effects are expected to be found also at other sites. \newline
\begin{small}
PACS: 96.40.Pq
\end{small}

\begin{keyword}
UHECR; composition; extensive air shower; atmosphere; shower maximum; atmospheric depth
\end{keyword}
\end{abstract}
\end{frontmatter}

\section{Introduction}
\label{sec:intro}

The number of charged particles in extensive air showers (EAS) as function of 
atmospheric depth, called longitudinal shower
profile, is closely related to the primary particle type and energy.  For a given energy, protons produce
showers that develop, on average, deeper in the atmosphere than showers of nuclei. The atmospheric depth at
which a shower exhibits its maximum of charged particles, $X_{\rm max}$, is well correlated 
with the mass of the primary particle. However, the stochastic nature of the individual 
particle production processes leads to large shower-to-shower fluctuations. On the other hand,
the size of the fluctuations depends on the mass number, too.
Therefore, both the mean depth of maximum and the width of the
$X_{\rm max}$ distribution hold important clues about the elemental composition of the primary shower
particle. 

The most direct method of measuring the longitudinal shower profile at shower energies above $10^{17}$\,eV
is the fluorescence light technique \cite{baltrusaitis}. Several existing
and upcoming experiments apply this technique, for example, HiRes~\cite{hires},
the Pierre Auger Observatory~\cite{PAO1,PAO2,PAO3}, Telescope Array~\cite{TA1,TA2},
and EUSO~\cite{EUSO1,EUSO2,EUSO3}. The fluorescence technique exploits
that charged particles traversing the atmosphere excite
nitrogen molecules. The de-excitation proceeds partially through fluorescence light emission,
mainly in the wavelength range between 300 and 400~nm.
It is expected and also indicated by direct measurements \cite{kakimoto} that the fluorescence 
yield is proportional to the local
ionization energy deposit of a shower. Therefore, the main advantage of this
observation technique is that the electromagnetic energy of the
EAS is obtained calorimetrically, i.e.~nearly independent of the primary cosmic ray composition. 
At the same time, the longitudinal shower profile can be reconstructed from the observed light profile.

The atmosphere plays a major role in fluorescence light based air shower experiments. 
On one hand, it serves as interaction target for
cosmic rays and as calorimeter in which the secondary particles deposit their energy via ionization 
losses. On the other hand, the atmosphere is the source of the detected fluorescence emission 
and also propagation medium for the light. Varying atmospheric conditions can influence the observed light signal
through all stages of the detection process, from shower development over  
fluorescence light production in the shower to light propagation to the detector.

This article is the first of a series of investigations of the importance of 
molecular atmospheric properties for the reconstruction of EAS.   
Within this article, we
shall restrict ourselves to geometrical effects implied by the shape and variation of the 
vertical atmospheric density profile. Forthcoming articles will address the dependence of the 
fluorescence yield and light propagation on
molecular atmospheric conditions.

All fluorescence detectors measure the light signal of showers as time traces with a certain angular
resolution. After reconstructing the shower core coordinates and direction, these time traces are mapped to a
geometric location in the atmosphere from which the observed light was emitted. In first approximation, 
this geometry reconstruction does not depend on atmospheric properties such as the local atmospheric density
at a given altitude. Considering the shower evolution, the situation is different. The interaction and
energy loss processes
of the shower particles depend mainly on the traversed atmospheric depth (i.e.~traversed column density) 
and not on the
particular geometry of the shower trajectory. Therefore it is the shower profile as function of atmospheric
depth that is most appropriate for characterizing air showers. Moreover it is this profile, and in
particular the depth of maximum, that is needed for composition studies. Varying atmospheric conditions
lead to the effect that, for example, the same geometrical height can correspond to different 
atmospheric depths. Of course, a misreconstruction of the shower axis could also lead to an incorrect 
determination of the shower depth profile.

In the following we shall study atmospheric effects on the
conversion of atmospheric depth, as used in shower simulations, to geometrical
altitude, as reconstructed from fluorescence measurements, and vice
versa. We will concentrate on the atmospheric conditions as found at the Pampa Amarilla, Argentina, at
the site of the southern Pierre Auger Observatory, where we  
have performed radio soundings of the molecular atmosphere for more than one year.

The structure of the article is the following. In Sec.~\ref{sec:atmos} the atmospheric data are
presented that will be used for the following discussion. The importance of using
realistic atmospheric profiles is demonstrated in Sec.~\ref{sec:longi}. First the conversion
between atmospheric depth and geometrical height is discussed for two actual
atmospheres. In the second part, the position of the shower maximum and its distribution is
presented taking into account measured variations of the atmosphere 
and averaged seasonal atmospheric models. 
A summary is given in Sec.~\ref{sec:summary}.

\section{Atmospheric conditions}
\label{sec:atmos}

The atmospheric depth associated with a given height plays a central role in EAS simulation and
reconstruction. The interaction
probability of a shower particle depends only on its traversed column depth, 
which can be expressed conveniently as difference between the atmospheric depths of the production and
interaction points. Similarly the conversion of height to atmospheric depth is needed for
mapping a reconstructed event geometry to the depth profile of a shower.  

The relation between atmospheric depth and height follows from the air density profile, whereas typically
the density profile of the US Standard Atmosphere 1976 (US-StdA)
is assumed~\cite{US-StdA1976}. Its atmospheric depth parameterization according to
J.~Linsley is implemented as the default profile in many Monte Carlo simulation programs like, for example,
CORSIKA~\cite{heckcorsika} and AIRES~\cite{sciuttoaires}. 
In a recent study the atmospheric conditions at the site of the Auger Observatory in
Argentina were investigated in detail~\cite{keilhauerICRC}.
The results important for the discussion of atmospheric profiles are summarized below.

\subsection{Measurement and analysis technique}
\label{subsec:technique}

Meteorological radio soundings were performed to measure altitude profiles of
temperature $T(h)$, air pressure $p(h)$, and relative humidity $u(h)$. 
The data were recorded using radiosondes~\cite{graw} launched on helium filled balloons. 
During ascent at least one data sample was taken every 8 seconds. Typically, the recording
frequency was even higher which gives an average height interval of 
about 20~m between two measurements. The balloons used reached altitudes
of about 20-25~km~a.s.l. 

During six measurement campaigns, 61 successful radio soundings were performed at the
Pampa Amarilla, Argentina, out of which 51 measurements
were done at night. Table~\ref{tab:launches} summarizes briefly the statistics of
the campaigns.
\begin{table}[htbp]
\caption{Statistics for the radio soundings performed at the Pampa Amarilla, Argentina.}
\begin{center}
\begin{tabular}{|c|c|c|}
\hline
\textbf{Date} & \textbf{Local Season} & \textbf{No.~of launches} \\ \hline
August 2002 & winter & 9 \\
November 2002 & spring & 9 \\
January / February 2003 & summer & 15 \\
April / May 2003 & autumn & 11 \\
July / August 2003 & winter & 8 \\
November 2003 & spring & 9 \\
\hline
\end{tabular}
\end{center}
\label{tab:launches}
\end{table}

The air density $\rho(h)$ is deduced from the measured data using
\begin{equation}
\rho(h) = \frac{p(h)\cdot M_\mathrm{m}}{R\cdot T(h)}\ ,
\label{eq:rho}
\end{equation}
where $R$ is the universal gas constant and $M_\mathrm{m}$ the molar mass of air in g/mol. 
The water vapor contribution to the molar mass is included on the basis of the measured relative humidity 
profiles~\cite{bodhaine}. For calculating the atmospheric depth, the following procedure is adopted. 
In the altitude region covered by balloon data, the height interval $\Delta h = h_{j+1} - h_j$ 
between two adjacent measurements
is sufficiently small, so that the local change of atmospheric depth is deduced from
\begin{equation}
\Delta X = \frac{\rho(h_j) + \rho(h_{j+1})}{2}\cdot(h_{j+1} - h_j)\ , \mathrm{~~where~~} h_{j+1} > h_j.
\end{equation}
The upper end of the measured profile is given by the altitude of balloon burst $h_\mathrm{b}$. There we
assume
\begin{eqnarray}
p(h_\mathrm{b})\ &=& \int_{h_{\rm b}}^\infty g(h)\cdot \rho(h)\ dh 
\ \approx\ g(h_{\rm b}) \int_{h_\mathrm{b}}^{\infty} \rho(h)\ dh \\
\Rightarrow X_\mathrm{b} &=& \frac{p(h_\mathrm{b})}{g(h_\mathrm{b})},
\end{eqnarray}
where $X_\mathrm{b}$ is the atmospheric depth at $h_{\rm b}$. The acceleration due to gravity
is denoted by $g(h_\mathrm{b})$ and calculated in dependence of 
the altitude for the geographic latitude of Malarg\"ue, Argentina. 
Thus, a full set of data describing the molecular aspects of the atmosphere
is obtained. The profiles for temperature, pressure, relative humidity,
density, and atmospheric depth represent the vertical structure of actual atmospheres at
the location of the southern Auger experiment with high vertical resolution. 

In a next step the derived atmospheric depth profiles are parameterized 
for applying them in the air shower simulation program CORSIKA. Following 
the functional form already used in CORSIKA \cite{heckcorsika} the depth 
profile is divided into four layers described by
\begin{equation}
X(h) = a_i + b_i\cdot \mathrm{e}^{-h/c_i}~, \mathrm{~with~} i=1,\ldots, 4.
\label{eq.gramparam}
\end{equation}
At very high altitude, between 100 and 112.8 km \cite{heckcorsika}, 
it is assumed that the atmospheric depth 
decreases linearly with height
\begin{equation}
X(h) = a_5 -b_5\frac{h}{c_5}.
\end{equation}

\subsection{Profiles at Pampa Amarilla, Argentina}
\label{subsec:pampa}

\begin{figure}[htbp]
\noindent
\begin{minipage}[c]{.48\linewidth}
\epsfig{file=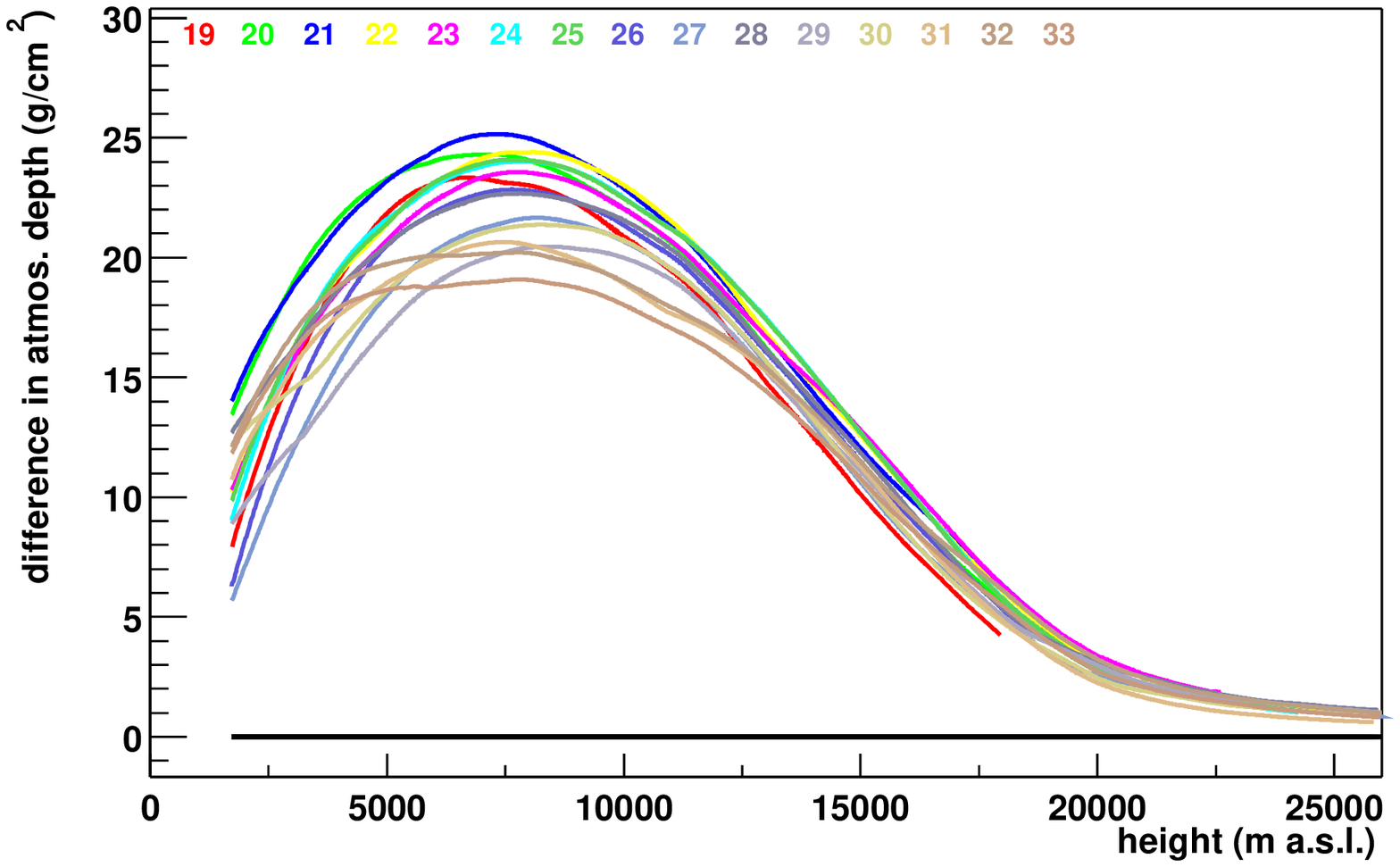,width=1.\linewidth}
\caption{Difference between the US-StdA depth profile and those measured during January/ February 2003,
near Malarg\"ue, Argen\-tina.}
\label{fig:diff_gram_summer}
\end{minipage}\hfill
\begin{minipage}[c]{.48\linewidth}
\epsfig{file=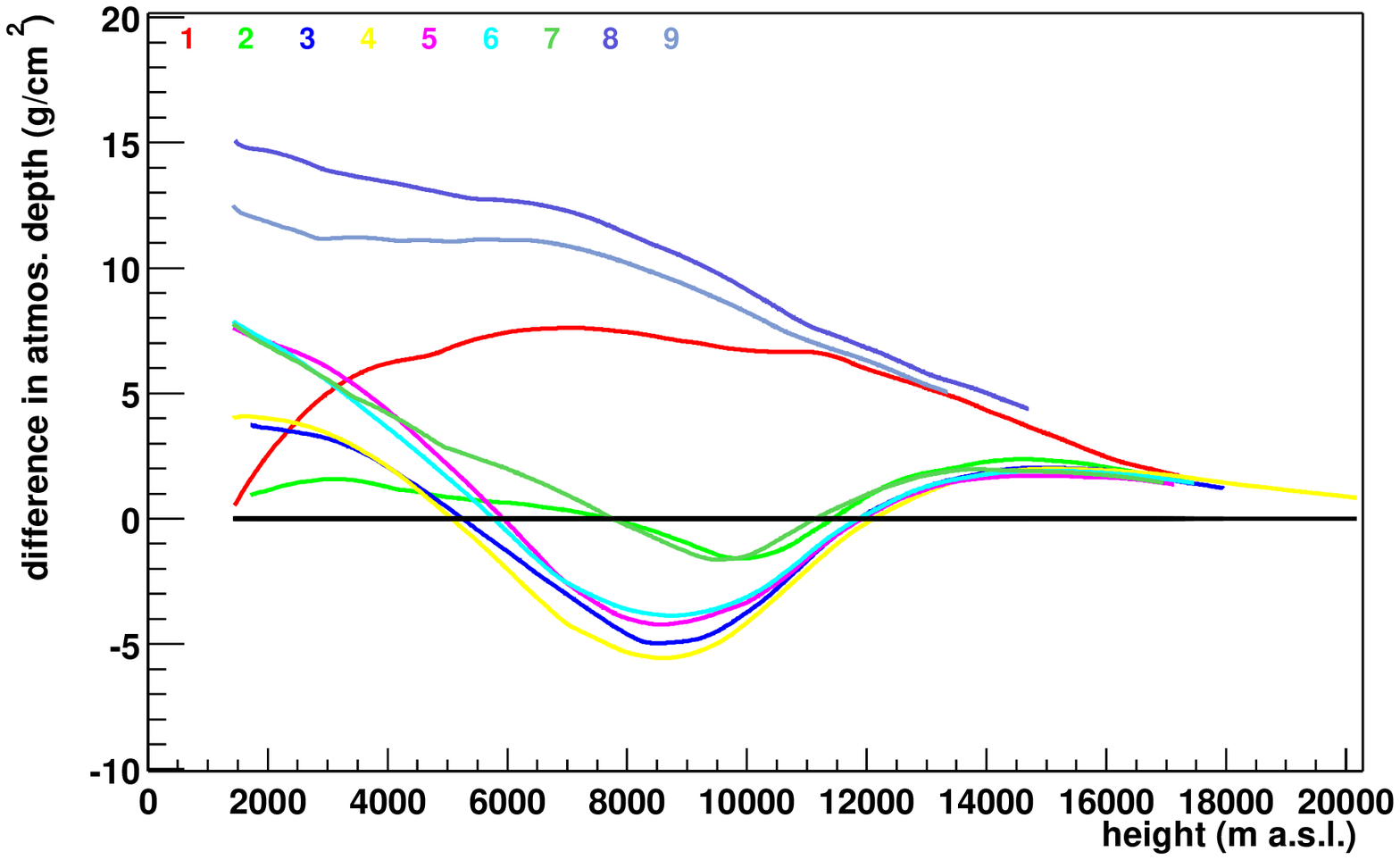,width=1.\linewidth}
\caption{Difference between the US-StdA depth profile and those measured during August 2002, near
Malarg\"ue, Argentina.} 
\label{fig:diff_gram_winter1}
\end{minipage}
\end{figure}
Atmospheric depth profiles (i.e. atmospheric depths as function of height) were derived
for each launch using the previously described procedure. 
Figs.~\ref{fig:diff_gram_summer} and  \ref{fig:diff_gram_winter1}
show the relative difference between the individual measured profiles and the US-StdA prediction. 
The launches are consecutively numbered for all campaigns and each line
corresponds to one launch. The variation of conditions of the 
atmosphere within summer was smallest. A representative data set is given in 
Fig.~\ref{fig:diff_gram_summer}, where the measurements taken during austral summer are shown. 
During winter, the conditions -- in particular those of air pressure and
atmospheric depth as the main influencing parameters on the EAS development
\cite{keilhauerICRC,wilczynskaICRC} -- differed quite strongly within several days,
see Fig.~\ref{fig:diff_gram_winter1}. 

\begin{figure}[bp]
\noindent
\begin{minipage}[c]{.48\linewidth}
\centering\epsfig{file=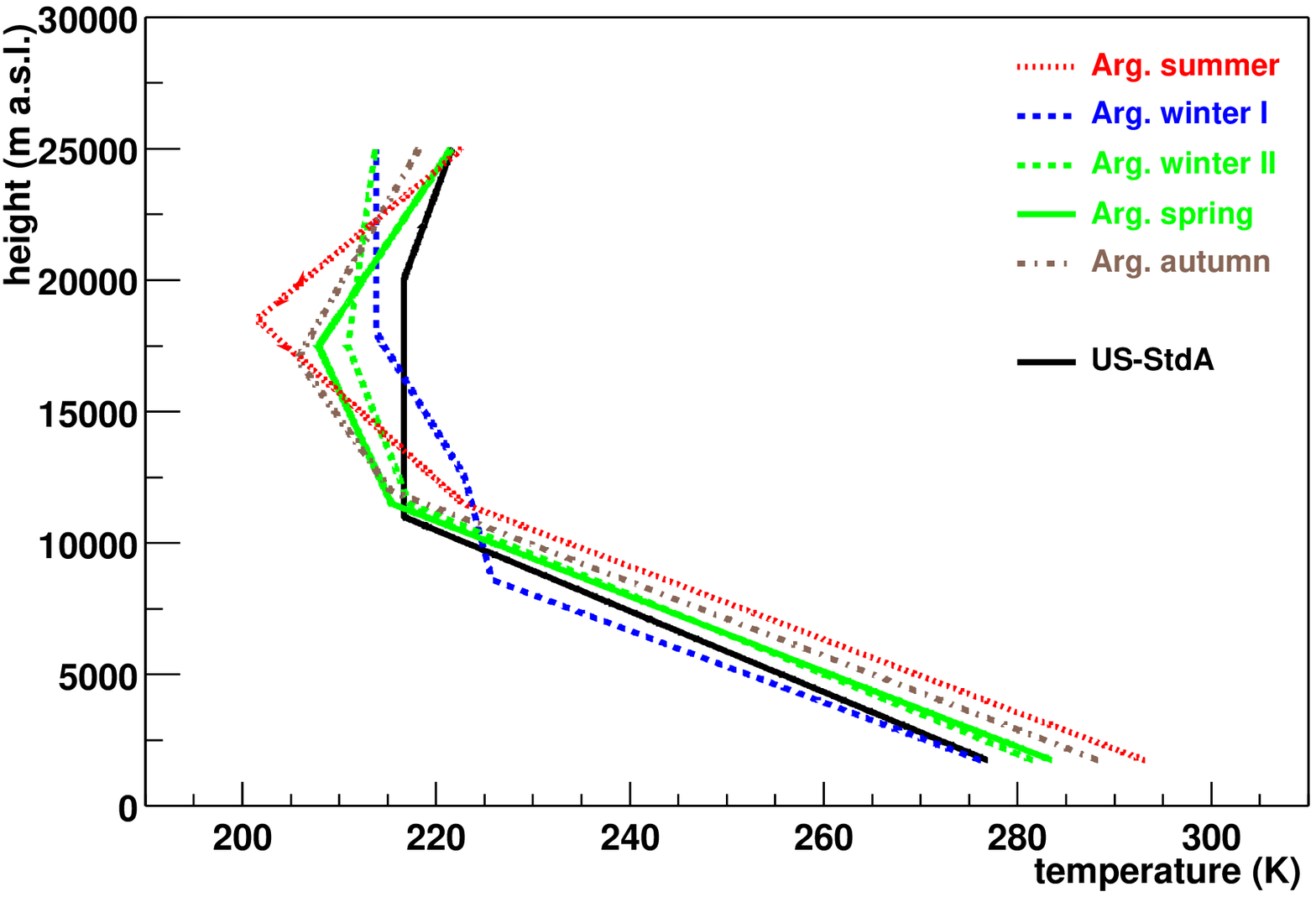,width=1.\linewidth}
\caption{Temperature profiles for averaged seasons at the Pampa Amarilla.}
\label{fig:temp_arg}
\end{minipage}\hfill
\begin{minipage}[c]{.48\linewidth}
\centering\epsfig{file=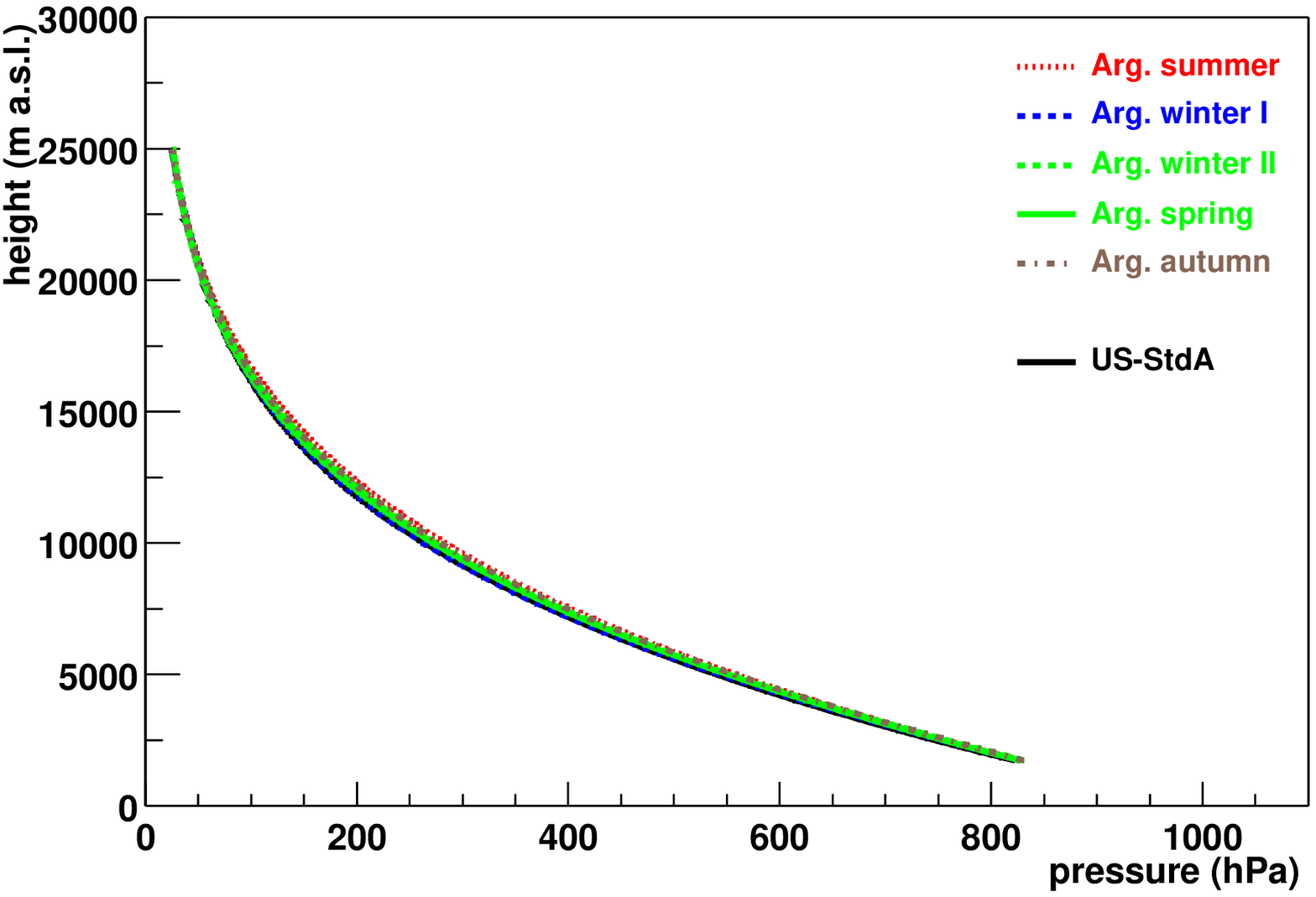,width=1.\linewidth}
\caption{Pressure profiles for averaged seasons at the Pampa Amarilla.}
\label{fig:pressure_arg}
\end{minipage}
\end{figure}

\begin{figure}[htbp]
\begin{center}
\epsfig{file=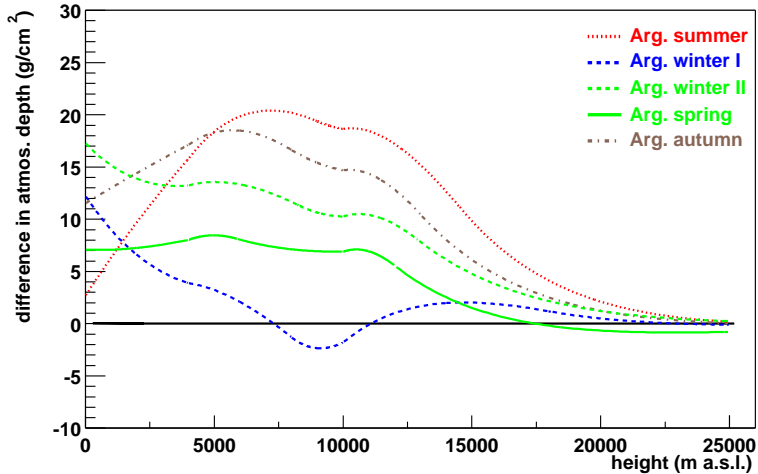,width=0.8\textwidth}
\end{center}
\caption{Differences in the average atmospheric 
depth profiles for the seasons in Argentina according to the US-StdA.}
\label{fig:diff_gram_arg}
\end{figure}

Atmospheric models were prepared averaging over each season, with the
exception of winter. For describing the winter conditions at Pampa Amarilla, two winter models are
required. Winter I reflects those situations when air pressure and atmospheric depth are
smaller than in the US-StdA at higher altitudes. The five obtained atmospheric models up
to 25~km~a.s.l.~are shown in Figs.~\ref{fig:temp_arg}, \ref{fig:pressure_arg}, and
\ref{fig:diff_gram_arg}, as temperature, pressure, and difference in atmospheric depth
profiles, respectively. 
These models are deduced from data of the first five
campaigns selecting only clear and calm nights with conditions that would allow 
operation of fluorescence detectors. For details concerning all radio sounding data
see~\cite{keilhauerFZKA}. The corresponding parameters for the atmospheric depth 
parameterizations are given in the Appendix.

\section{Atmospheric Influences on the Reconstruction of Extensive Air Shower Profiles}
\label{sec:longi}

Fluorescence telescopes detect the longitudinal shower development within a fixed field of
view. The visible height range depends on the distance of the EAS to the telescope. For
analyzing simulated shower profiles in the geometrical frame of telescopes, these profiles have to be
transformed from a description based on vertical atmospheric depth $X_\mathrm{vert}$ to geometrical
height $h$. However, the development of an EAS depends on the slant atmospheric
depth $X$ as the amount of traversed matter which is given by $X = X_\mathrm{vert}/\mathrm{cos}\theta$
for zenith angle $\theta$ of the EAS less than 60$^\circ$. Due to this fact, the following
discussion depends strongly on the zenith angle of an EAS. For vertical
showers, the geometrical conversion effect is smallest.  

From the physical point of view, an EAS develops according to the amount of traversed
air. Starting with the first interaction of the primary cosmic ray particle, the
cascade of secondary particles increases in number of particles as well as in energy
deposited in air and fluorescence emission. At the shower maximum, energy losses begin to dominate
particle production and the number of particles decreases while propagating further in the 
atmosphere. The position of the shower maximum in terms of
atmospheric depth is closely correlated to the type of the primary particle and is used to
identify the cosmic ray for a given primary energy $E_0$. Fluorescence telescopes however,
observe EAS in terms of geometrical height $h$. Thus, observed profiles have to be
converted into atmospheric depth related profiles for an interpretation of the event. As
mentioned above, usually the profile of the US-StdA is applied for this purpose. 

\subsection{Importance of using realistic atmospheric profiles}
\label{subsec:importance}

For this study, 100 iron and 200 proton induced EAS have been simulated using CORSIKA 
with the hadronic interaction model QGSJET01
\cite{kalmykov}. In the following we will use only the average shower profiles calculated from 
the set of simulations. All showers were generated using the US-StdA and the shower profile is 
tabulated as function of atmospheric depth. It is sufficient to simulate showers in one atmosphere as the 
physical development of a shower is only slightly affected by varying 
atmospheric conditions in terms of changing particle interaction and decay probabilities due to air density 
distributions. Differences in the observable shower 
profile as function of altitude are adequately considered by using realistic 
atmospheric profiles for the conversion of atmospheric depth to geometrical altitude.

\begin{figure}[htbp]
\begin{center}
\epsfig{file=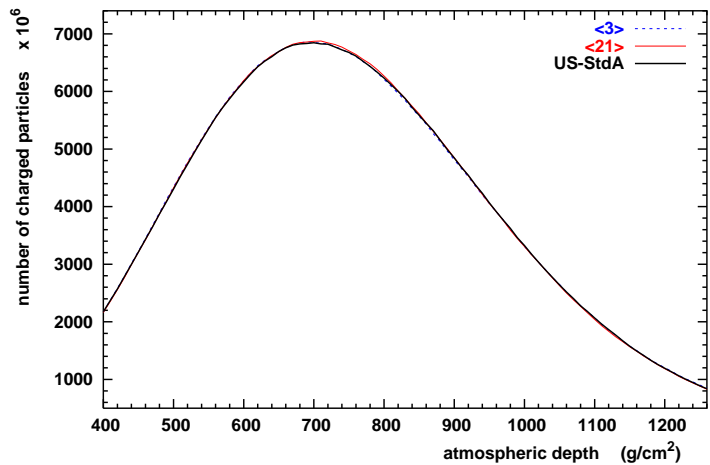,width=0.7\textwidth}
\end{center}
\caption{
Mean profiles of 100 iron induced showers with $10^{19}$~eV and 60$^\circ$ incident angle in different atmospheres. 
The simulations were done by implementing the corresponding atmospheric profiles in CORSIKA.
} 
\label{fig:diff_shower}
\end{figure}

This simplified treatment is justified since  
the shower development depends only very weakly on the used atmospheric depth profile due to different atmospheres. 
Fig.~\ref{fig:diff_shower} shows the mean shower profiles of 
iron induced showers with 10$^{19}$~eV and 60$^\circ$ zenith angle. Using two
extreme Argentine atmospheres and the US-StdA sets of 100 showers each were simulated. 
Only very small differences in the shower profile can be observed. 

\begin{figure}[tbp]
\noindent
\begin{minipage}[c]{.48\linewidth}
\centering\epsfig{file=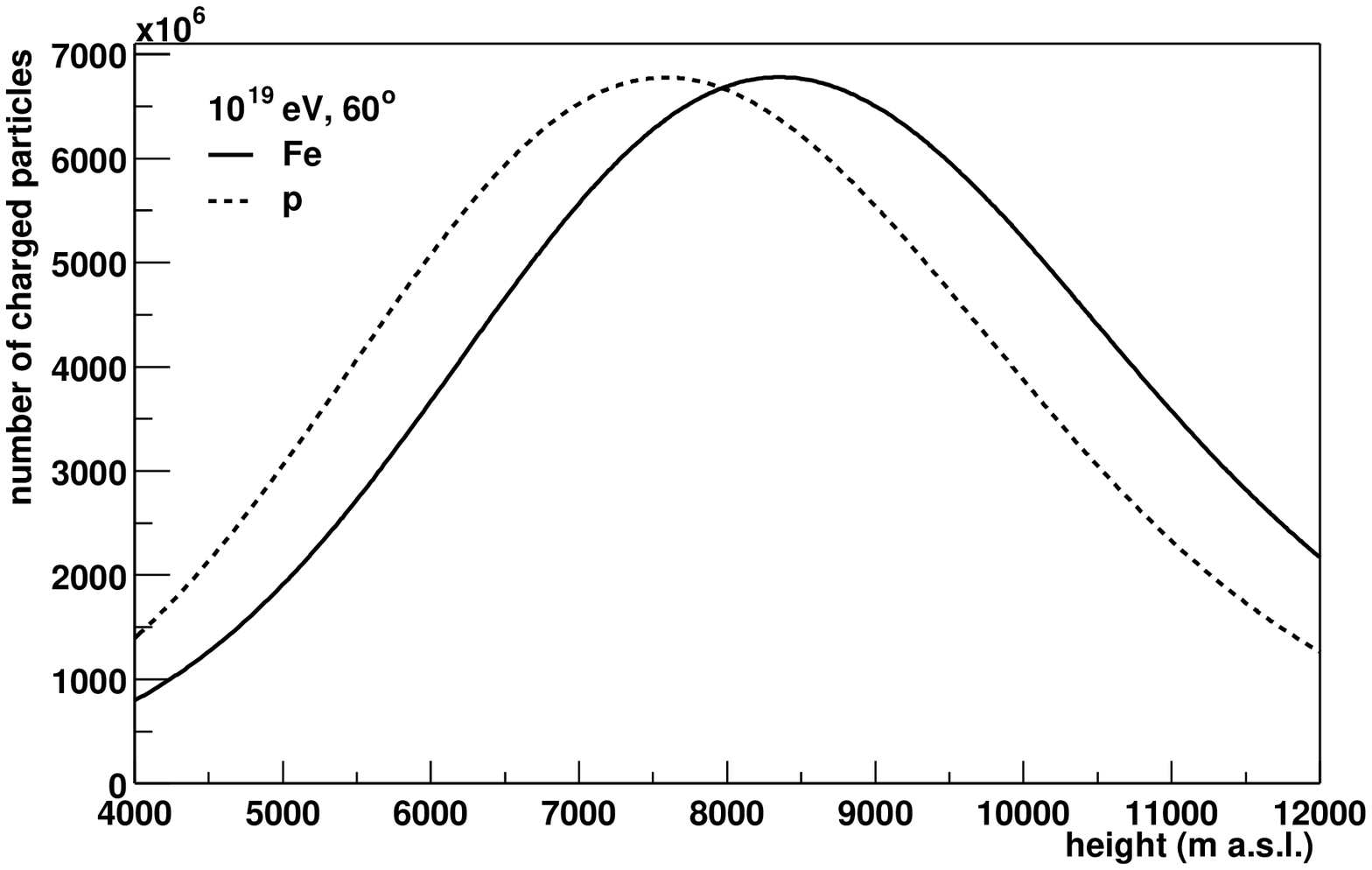,width=1.\linewidth}
\caption{Profiles of charged particles for p- and Fe-induced EAS in US-StdA with
10$^{19}$~eV and 60$^\circ$ inclination.}
\label{fig:particle_pfe_us}
\end{minipage}\hfill
\begin{minipage}[c]{.48\linewidth}
\centering\epsfig{file=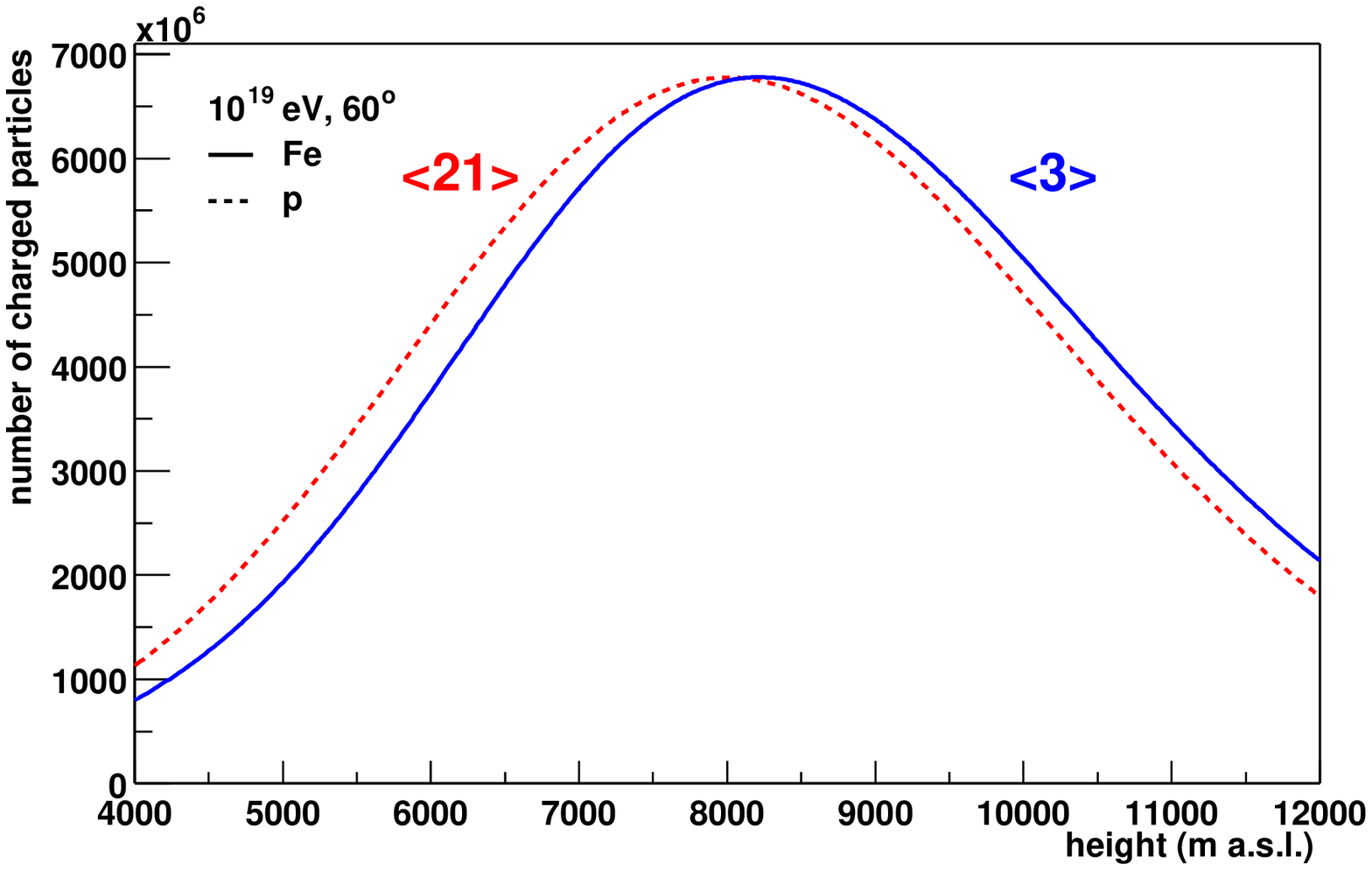,width=1.\linewidth}
\caption{Profiles of charged particles for p-induced EAS in Argentine summer, launch $<$21$>$, and Fe-induced EAS in
Argentine winter, launch $<$3$>$, both with 10$^{19}$~eV and 60$^\circ$ inclination.}
\label{fig:particle_ps_few}
\end{minipage}
\end{figure}

In Fig.~\ref{fig:particle_pfe_us},
the comparison of charged particles of p- and Fe-induced showers with $E_0 =
10^{19}$ eV and $\theta = 60^\circ$ in the US-StdA can be seen. For this, the underlying
simulated EAS in terms of atmospheric depth have been transformed to geometrical height
using the parameterization for the US-StdA. Iron induced showers
develop earlier and the average position of the shower maximum is reached for the given
conditions at $8.4\;$km. Proton induced showers penetrate deeper in the atmosphere and the
position of the shower maximum is at $7.6\;$km. The EAS profiles are clearly separated and a
discrimination by fluorescence telescope detection is expected. 

However, actual atmospheric conditions differ from the US-StdA as shown in Sec.~\ref{sec:atmos}. 
For demonstrating the importance of using a realistic atmospheric profile, we show
the same averaged showers in Fig.~\ref{fig:particle_ps_few}, as they would be observed under 
two extreme cases of measured Argentine conditions.
The first case is a night during winter, recorded as radio sounding \asc{3},
and the second case is a night in summer, launch \asc{21}. The accordant atmospheric
depths can be seen in Fig.~\ref{fig:diff_gram_winter1} and \ref{fig:diff_gram_summer},
respectively. This corresponds to the assumption that the deeper penetrating p-induced shower 
would have been measured in austral summer
leading to a shift of longitudinal development towards higher altitudes. The
Fe-induced shower is assumed to occur during winter. For these conditions, the positions
of the shower maxima of the averaged proton and iron showers are very close together. 
The shower maximum of the Fe-induced shower
in winter is at 8.2~km, that of the p-induced shower in summer is at 8.0~km. Thus the
profiles are hardly distinguishable.

\begin{figure}[tbp]
\noindent
\begin{minipage}[c]{.48\linewidth}
\centering\epsfig{file=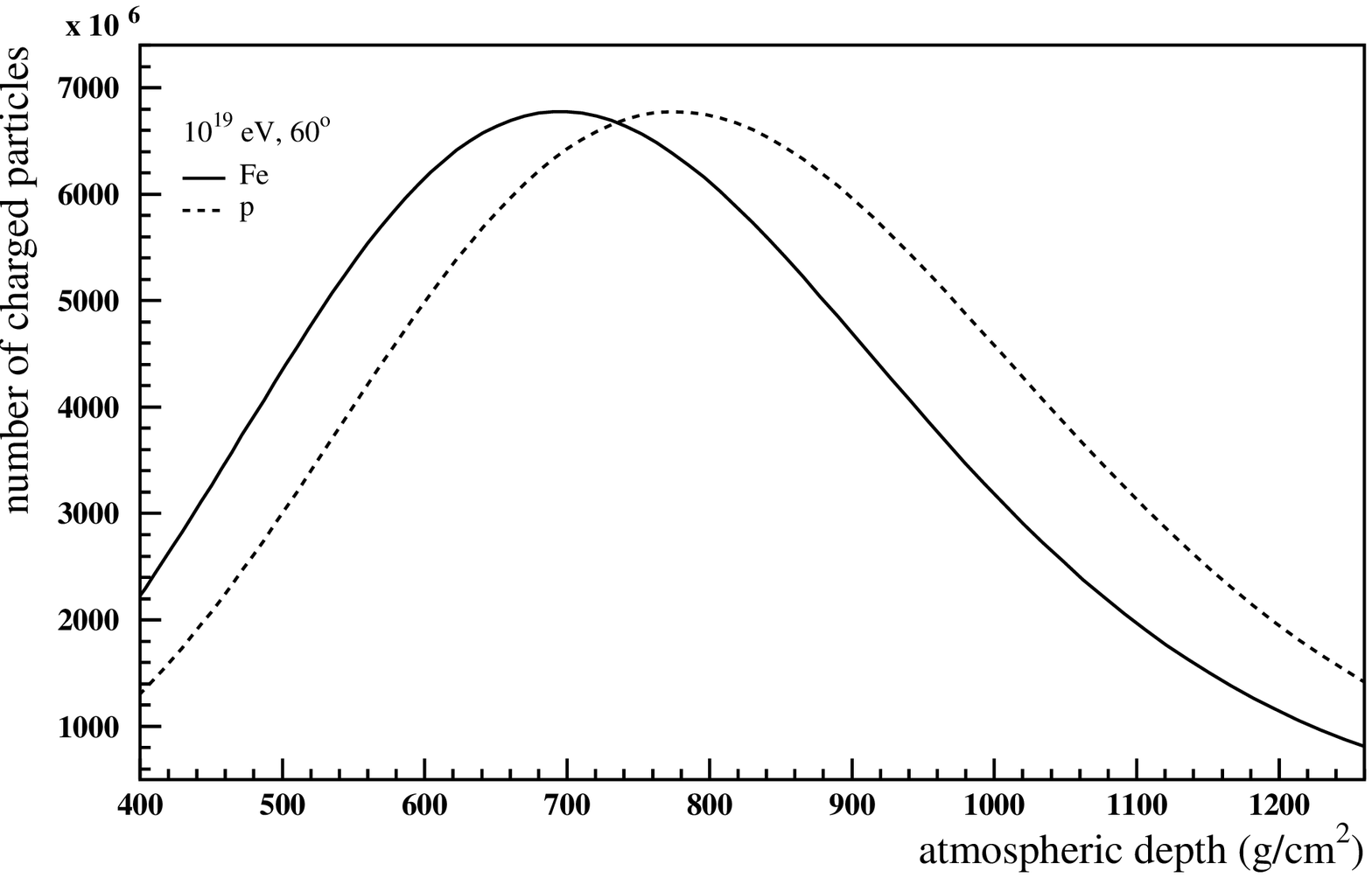,width=1.\linewidth}
\caption{Profiles of charged particles for p- and Fe-induced EAS in US-StdA with
10$^{19}$~eV and 60$^\circ$ inclination.}
\label{fig:gram_particle_fep_us}
\end{minipage}\hfill
\begin{minipage}[c]{.48\linewidth}
\centering\epsfig{file=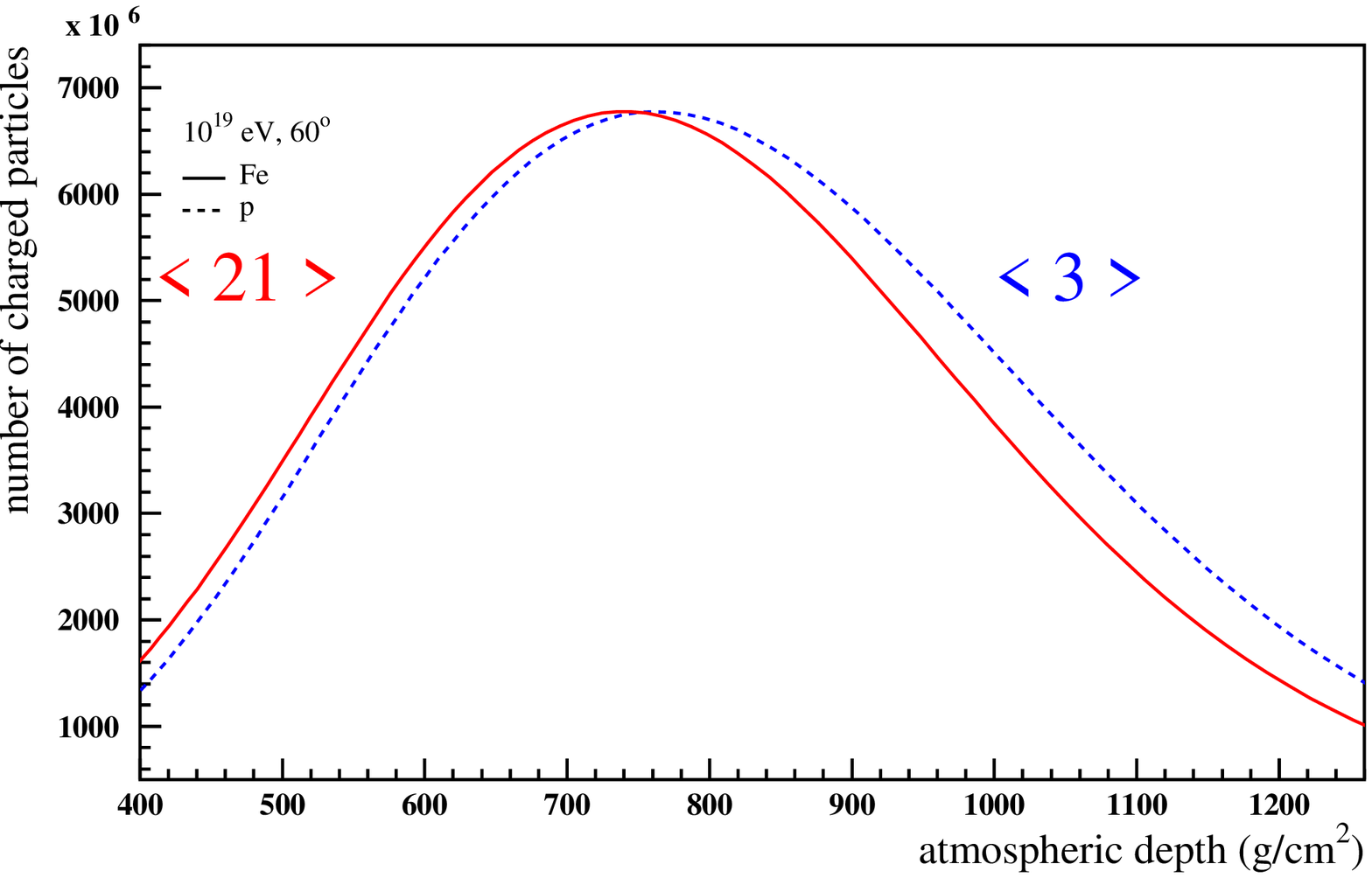,width=1.\linewidth}
\caption{Profiles of charged particles for Fe-induced EAS in Argentine summer, launch $<$21$>$, and p-induced EAS in
Argentine winter, launch $<$3$>$, both with 10$^{19}$~eV and 60$^\circ$ inclination.}
\label{fig:gram_particle_fe21_p3}
\end{minipage}
\end{figure}

Next we consider the number of charged particles in EAS
as a function of atmospheric depth $X$. The longitudinal profiles for the simulated p- and Fe-induced 
EAS in the US-StdA are
plotted in Fig.~\ref{fig:gram_particle_fep_us}. The shown range of the slant atmospheric
depth is nearly the same as the range in geometrical altitude given in
Figs.~\ref{fig:particle_pfe_us} and \ref{fig:particle_ps_few}. Again, the profiles are clearly
distinguishable and the position of the shower maximum for the proton case is at
$774\;$g/cm$^2$ and for the iron shower at $690\;$g/cm$^2$.  Adopting the
reconstruction point of view, now the EAS profiles in terms of geometrical height are
taken as given. The question is how large are the shifts
of the shower profiles of Fig.~\ref{fig:particle_pfe_us} in $X$, if the profiles of actually measured
atmospheres are used. Applying atmosphere \asc{21} to iron induced EAS and \asc{3} to
p-induced showers, the resulting charged particle profiles are plotted in
Fig.~\ref{fig:gram_particle_fe21_p3}. The position of the shower maximum of the iron shower in atmosphere
\asc{21} would be reconstructed to $735\;$g/cm$^2$ and for the p-induced shower to $761\;$g/cm$^2$. 
Relative to US-StdA this corresponds to a shift of about $+45$~g/cm$^2$ and $-13$~g/cm$^2$ in slant depth 
for the iron and proton showers, respectively.

\begin{figure}[htbp]
\begin{center}
\epsfig{file=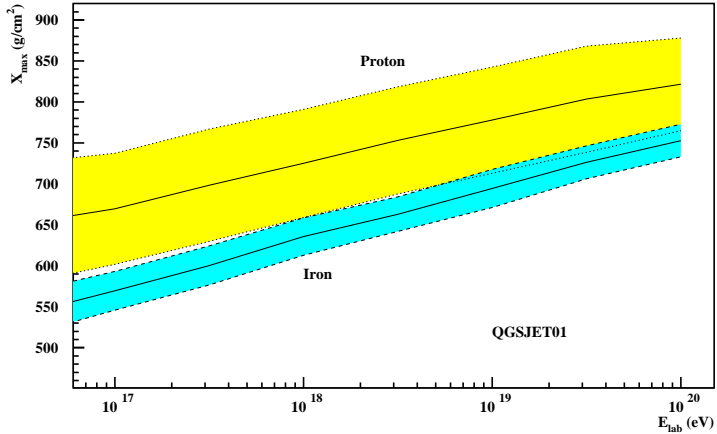,width=0.7\textwidth}
\end{center}
\caption{Mean value and fluctuations of the depth of shower maximum for increasing $E_0$. 
The showers were simulated with CORSIKA 
applying the QGSJET01 hadronic interaction model~\protect\cite{kalmykov}. For p-induced showers 500
simulations were performed and for Fe-induced 200. The bands denote one standard deviation.}
\label{fig:xmax_baender_p_fe}
\end{figure}

\begin{figure}[htbp]
\begin{center}
\epsfig{file=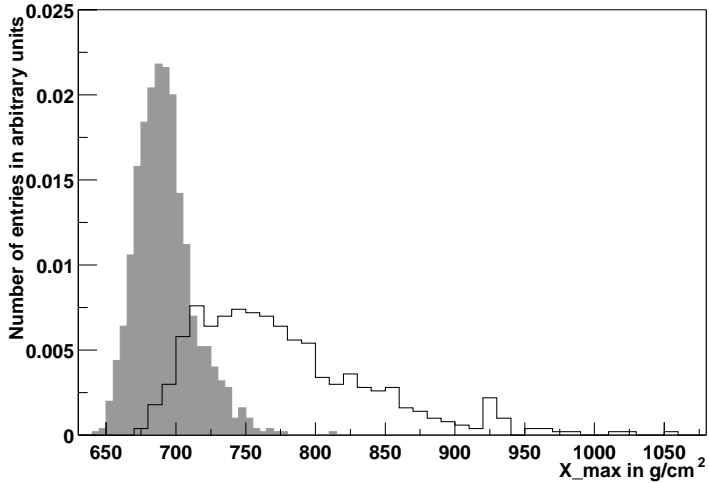,width=0.75\textwidth}
\end{center}
\caption{$X_{\rm max}$ distribution for 1000 Fe-induced (shaded histogram) and 
500 p-induced (white), 10$^{19}$~eV showers given in slant depth.}
\label{fig:xmax_us_Fep}
\end{figure}

Not only the average position of the shower maximum is indicating the type of the primary
particle. Also the width of the distribution of this position for a large number of EAS is
systematically different for proton and iron induced showers.
The general relation between shower maximum and $E_0$ of EAS in the US-StdA can be seen in
Fig.~\ref{fig:xmax_baender_p_fe}. 
As a concrete example, we show in Fig.~\ref{fig:xmax_us_Fep} the $X_{\rm max}$ distributions of 
proton and iron induced
showers at $10^{19}\;$eV for US-StdA. The fluctuations are much
larger and more asymmetric for proton induced showers ($\sigma_{\rm p} \approx$ 60~g/cm$^2$) 
than for iron induced EAS ($\sigma_{\rm Fe}\approx$ 20~g/cm$^2$).
Some characteristics of these distributions are listed in
Table~\ref{tab:histo_slant_us_pFe}.
\begin{table}[htbp]
\caption{Statistics for the $X_\mathrm{max}$ distributions of Figure
\ref{fig:xmax_us_Fep}.}
\begin{center}
\begin{tabular}{|c||c|c|c|c|c|}
\hline
 & $\mathbf{N_\mathrm{entry}}$ & \textbf{Mean} & \textbf{RMS} & \textbf{Minimal Value} &
\textbf{Maximal Value} \\
 & & (g/cm$^2$) & (g/cm$^2$) & (g/cm$^2$) & (g/cm$^2$) \\ \hline\hline
\textbf{Fe-ind.} & 1000 & 692 & 20.9 & 644 & 811 \\
\textbf{p-ind.} & 500 & 778 & 64.7 & 671 & 1056 \\
\hline
\end{tabular}
\end{center}
\label{tab:histo_slant_us_pFe}
\end{table}

The information of the width of
the distribution of the maximum position of EAS is often used in experiments for
deducing the chemical composition of primary cosmic rays from the
data~\cite{cronin,watson}. 
Since for measuring the $X_{\rm max}$ distribution, many EAS events have to be taken
into account, however, the atmospheric conditions are different for each event. 
As demonstrated earlier, the use of the US-Std atmosphere can cause a shift of  
up to $\sim 40\;$g/cm$^2$ for $60^\circ$ showers for atmospheric 
profiles measured at Pampa Amarilla, Argentina. If the local atmosphere above a 
fluorescence detector changes as much as considered in the extreme examples above on a day-by-day basis,
such shifts would considerably broaden 
the $X_{\rm max}$ distribution of heavier elements.  A reconstruction based on a fixed 
atmospheric profile
would lead to a systematic bias towards lighter mass numbers.   
In the next section, we shall study in detail the influence of different, measured atmospheric 
conditions on the mean and width of the distribution of the shower maximum.

\subsection{Application of averaged seasonal atmospheres for the Pampa Amarilla}
\label{subsec:arg_appli}

\begin{figure}[htbp]
\noindent
\begin{minipage}[c]{.49\linewidth}
\centering\epsfig{file=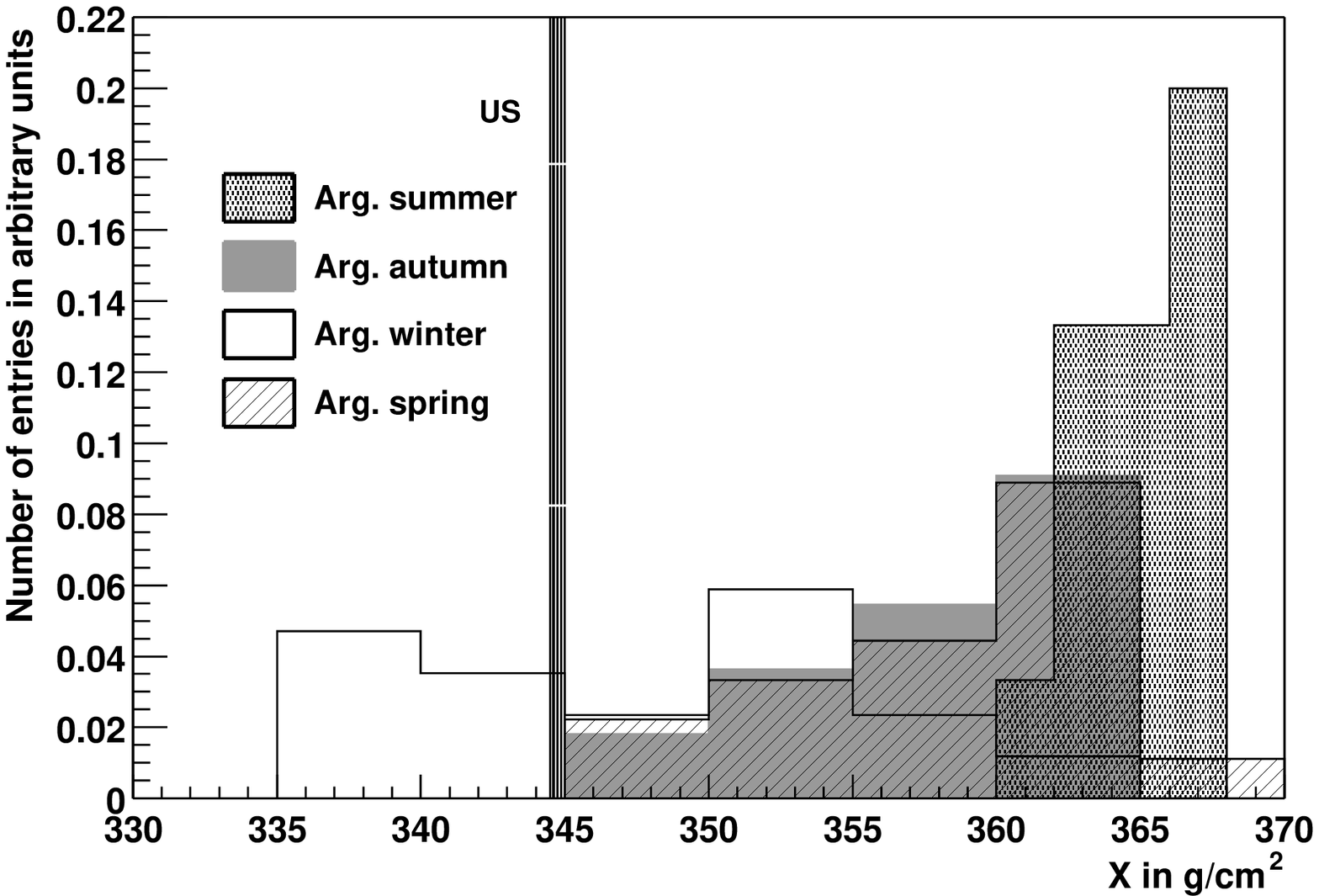,width=1.\linewidth}
\caption{$X$ distribution at 8400~m~a.s.l.~for the individual profiles measured in
Argentina divided into groups by season.}
\label{fig:gram_histo_8400}
\end{minipage}\hfill
\begin{minipage}[c]{.49\linewidth}
\centering\epsfig{file=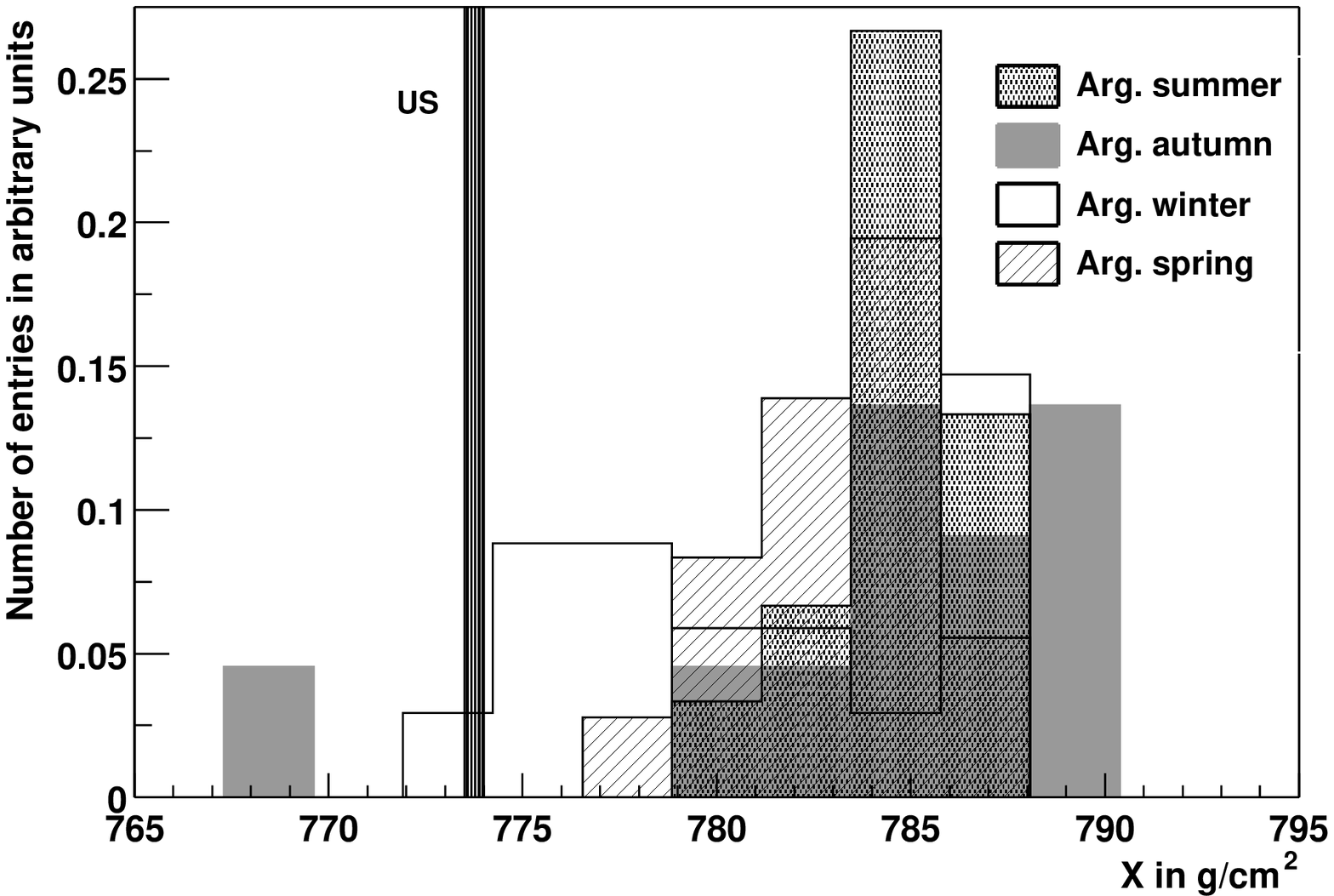,width=1.\linewidth}
\caption{X distribution at 2400~m~a.s.l.~for the individual profiles measured in Argentina divided into groups by season.}
\label{fig:gram_histo_2400}
\end{minipage}
\end{figure}

First we use our measurements to analyze the atmospheric variations in each season. 
For two representative altitudes,
the corresponding vertical atmospheric depth values obtained with the radio soundings are
calculated. In Fig.~\ref{fig:gram_histo_8400}, the distributions of the
atmospheric depth for the geometrical height of 8400~m~a.s.l.\ are plotted for different seasons.
The height of
8400~m~a.s.l.\ is chosen because it corresponds approximately to the altitude where a
Fe-ind.\ EAS with 10$^{19}$~eV and 60$^\circ$ inclination reaches its maximum. The
statistics for all four seasons is given in Table~\ref{tab:stat_8400}, admittedly in slant
depth underlying a 60$^\circ$ zenith angle. Argentine Summer is
marked by a small distribution in contrast to winter. The total width of the distribution
for all seasons is 8.2~g/cm$^2$ in vertical depth or 16.4~g/cm$^2$ in slant depth for the
given case. The second example is specified for 2400~m~a.s.l., being about the
altitude of the shower maximum for a p-ind.~EAS with 10$^{19}$~eV and vertical
incidence. The histogram is shown in Fig.~\ref{fig:gram_histo_2400} and the statistics is given in 
Table~\ref{tab:stat_2400}. Again, Argentine summer shows only small variations and at this
altitude, Argentine autumn spreads most. However, the width of the distribution for all
seasons at this height is only about half the value than for the higher altitude.
\begin{table}[htbp]
\noindent
\begin{minipage}[c]{.48\textwidth}
\caption{Statistics for the slant depth distributions at 8400~m~a.s.l. for the individual
profiles in Argentina and $60^\circ$ zenith angle.}
\label{tab:stat_8400}
\begin{center}
\begin{tabular}{|c||c|c|c|}
\hline
 & $\mathbf{N_\mathrm{entry}}$ & \textbf{Mean} & \textbf{RMS} \\
 & & (g/cm$^2$) & (g/cm$^2$)  \\ \hline\hline
\textbf{all}    & 61 & 714 & 16.4 \\ \hline
\textbf{Summer} & 15 & 730 & 3.4 \\
\textbf{Winter} & 17 & 696 & 15.2 \\
\textbf{Spring} & 18 & 716 & 10.6 \\
\textbf{Autumn} & 11 & 718 & 9.2 \\
\hline
\end{tabular}
\end{center}
\end{minipage}\hfill
\begin{minipage}[c]{.48\textwidth}
\caption{Statistics for the slant depth distributions at 2400~m~a.s.l.\ for the individual
profiles in Argentina and $0^\circ$ zenith angle.}
\label{tab:stat_2400}
\begin{center}
\begin{tabular}{|c||c|c|c|}
\hline
 & $\mathbf{N_\mathrm{entry}}$ & \textbf{Mean} & \textbf{RMS} \\
 & & (g/cm$^2$) & (g/cm$^2$)  \\ \hline\hline
\textbf{all} & 61 & 783 & 4.0 \\ \hline
\textbf{Summer} & 15 & 785 & 2.0 \\
\textbf{Winter} & 17 & 781 & 4.7 \\
\textbf{Spring} & 18 & 783 & 2.2 \\
\textbf{Autumn} & 11 & 783 & 5.5 \\
\hline
\end{tabular}
\end{center}
\end{minipage}\hfill
\end{table}

The intrinsic variation of the depth of maximum of proton induced showers is so large that the 
broadening effect due to the atmosphere becomes negligible. 
However for iron induced showers, a considerable effect due to atmospheric variations 
is expected. Additionally, it is obvious that the
influence is more important for inclined showers since the slant depth has to be
considered for the longitudinal shower development. The difference between US-StdA
and the actual atmosphere, as for instance given in Fig.~\ref{fig:diff_gram_arg}, has to be divided by
$\mathrm{cos}\theta$, with $\theta$ being the EAS zenith angle. The resulting value
represents the shift of the shower profile at each altitude.

The broadening effect of the $X_\mathrm{max}$ distribution due to all averaged Argentine
seasons and also the shift of the average position of $X_\mathrm{max}$ is demonstrated
in Tables~\ref{tab:histo_slant60_Fe} and
\ref{tab:histo_slant60_p} for 10$^{19}$~eV showers of 60$^\circ$ zenith angle.
\begin{table}[htbp]
\caption{Statistics for $X_{max}$ distributions for Fe-induced showers with
60$^\circ$ incidence in slant depth.}
\begin{center}
\begin{tabular}{|c||c|c|c|c|c|}
\hline
& $\mathbf{N_\mathrm{entry}}$ & \textbf{Mean} & \textbf{RMS} & \textbf{Min.~Value} &
\textbf{Max.~Value} \\
& & (g/cm$^2$) & (g/cm$^2$) & (g/cm$^2$) & (g/cm$^2$) \\ \hline\hline
\textbf{US-StdA} & 1000 & 692 & 20.9 & 644 & 811 \\
\textbf{Arg. averaged} & 5000 & 713 & 26.1 & 658 & 851 \\ \hline
\textbf{Arg. Summer} & 1000 & 732 & 21.2 & 683 & 851 \\
\textbf{Arg. Winter I} & 1000 & 688 & 21.4 & 640 & 811 \\
\textbf{Arg. Winter II} & 1000 & 715 & 21.3 & 666 & 836 \\
\textbf{Arg. Spring} & 1000 & 706 & 21.0 & 658 & 825 \\
\textbf{Arg. Autumn} & 1000 & 725 & 21.4 & 675 & 846 \\
\hline
\end{tabular}
\end{center}
\label{tab:histo_slant60_Fe}
\end{table}
\begin{table}[htbp]
\caption{Statistics for $X_{max}$ distributions for p-induced showers with
60$^\circ$ incidence in slant depth.}
\begin{center}
\begin{tabular}{|c||c|c|c|c|c|}
\hline
 & $\mathbf{N_\mathrm{entry}}$ & \textbf{Mean} & \textbf{RMS} & \textbf{Min.~Value} &
\textbf{Max.~Value} \\
 & & (g/cm$^2$) & (g/cm$^2$) & (g/cm$^2$) & (g/cm$^2$) \\ \hline\hline
\textbf{US-StdA} & 500 & 778 & 64.7 & 671 & 1056 \\
\textbf{Arg. averaged} & 2500 & 800 & 67.3 & 667 & 1094 \\ \hline
\textbf{Arg. Summer} & 500 & 818 & 64.7 & 710 & 1094 \\
\textbf{Arg. Winter I} & 500 & 777 & 66.7 & 667 & 1062 \\
\textbf{Arg. Winter II} & 500 & 802 & 65.7 & 693 & 1083 \\
\textbf{Arg. Spring} & 500 & 792 & 65.1 & 685 & 1073 \\
\textbf{Arg. Autumn} & 500 & 812 & 65.9 & 703 & 1093 \\
\hline
\end{tabular}
\end{center}
\label{tab:histo_slant60_p}
\end{table}
Regarding all seasons, the average shift of the position of the shower maximum is 21~g/cm$^2$ for
the Fe-ind.~shower and 22~g/cm$^2$ for p-ind. These number could also be extracted from
Fig.~\ref{fig:diff_gram_arg}. The average difference in atmospheric depth $\Delta X$ according to the
US-StdA is about 10~g/cm$^2$ at $\approx$ 8~km~a.s.l. For these examples of 60$^\circ$
inclined showers, it results into a shift in slant depth of $\Delta X/\mathrm{cos}\theta
\approx $ 20~g/cm$^2$. For each season, the number can be obtained in the same way. The
numbers for the variation of the $X_\mathrm{max}$ distribution indicate only small
atmospheric effects within each season. However for the average of all seasons, the
atmospheric conditions affect the variation especially in the case of iron. The
distribution becomes wider for about 25\% for iron but only 4\% for p-ind.~showers
compared to the distribution in the US-StdA.

For the sake of completeness it should be mentioned that the  
shift of the average $X_{\rm max}$ position is
correspondingly half of the values above for vertical showers. The width of the
distribution is unaffected. The numerical values for the 
different seasons can be taken from Fig.~\ref{fig:diff_gram_arg}.
In this case the broadening is negligible for all primaries considered here.

\section{Summary and Conclusion}
\label{sec:summary}

The importance of using realistic atmospheric depth profiles, strictly speaking air density profiles,
for reconstructing the longitudinal shower development has been investigated. The depth of
shower maximum has been considered in detail since it is often used for identifying the
type of the primary particle. Applying atmospheres measured at site of the southern Pierre Auger 
experiment, two main effects are observed with regard to using the US
standard atmosphere parameterization for data analysis and simulation studies.
First of all, atmospheric conditions differing from the
US-StdA can lead to a significant, systematic shift of the position of the shower maximum. 
Secondly, the distribution of the depth of maximum due to
intrinsic shower fluctuations is broadened by temporal variations of the
atmospheric conditions.

The importance of atmospheric variations depends on the shower angle and primary particle. 
The more inclined an air shower is the more important is the detailed knowledge 
of the atmospheric profile. For
vertical EAS, the influence of atmospheric profile variations 
can be nearly neglected, but for incidence angles larger
than 40$^\circ$, the use of the US standard atmosphere biases the interpretation of extensive air
showers. Showers induced by heavy particles are more sensitive to atmospheric profile deviations
than that of light primaries.

\begin{figure}[htbp]
\begin{center}
\epsfig{file=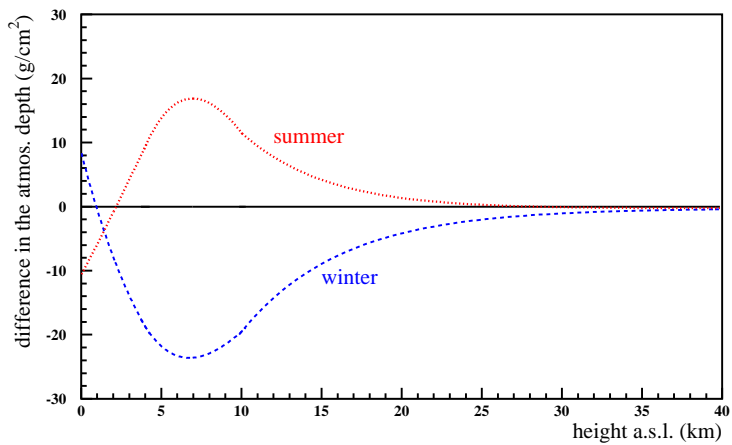,width=0.7\textwidth}
\end{center}
\caption{Difference in the atmospheric depth in g/cm$^2$ from the Stuttgart,
Germany, summer~/~winter atmosphere to the US-StdA \cite{ulrich}.} 
\label{fig:diff_gram_german}
\end{figure}

This
investigation is based on meteorological data obtained at the Pampa Amarilla, Argentina,
where the southern part of the Pierre Auger Observatory is situated. However, it is expected that
the atmospheric density profiles are as important for other fluorescence air shower experiments 
as for the Auger experiment. This is supported by Fig.~\ref{fig:diff_gram_german}, in which
deviations of the atmospheric depth
profiles relative to the US-StdA  are shown for two extreme profiles measured in Germany. The curve
for summer is very similar to Argentine summer and the deviation in winter even exceeds that
of the Argentine winter.

Seasonal atmospheric depth profiles have been developed for the southern Auger detector. With the
exception of Argentine winter, each season can be described by a typical shape of the atmospheric 
profile. Although the individual measurements correlate for the other seasons well with 
seasonally averaged atmospheric conditions, day-to-day variations are significant. Currently our
investigations are limited  by the statistics of radio soundings. We will continue to measure 
the molecular atmosphere at the Pampa Amarilla and plan to develop atmospheric models for  
various time scales. 

Forthcoming articles of this series will study the impacts of using realistic temperature and pressure
profiles on the fluorescence yield and on the transmission due to Rayleigh scattering of the
fluorescence light. In both cases, a wavelength dependent analysis has to be
performed. Similar analyses have also be done for the impact on the atmospheric Cherenkov
technique~\cite{bernloehr}. Only minor effects -- in addition to corrections for the
variations of the atmospheric ground pressure -- are expected for the observation of EAS
using particle detectors at ground. 

\section*{Acknowledgment}

The authors thank their colleagues from the Pierre Auger Collaboration, in particular
B.~Wilczy\'nska, H.~Wilczy\'nski, J.~A.~J.~Matthews, M.~Roberts, and V.~Rizi for many
inspiring and fruitful discussions. The help and support of D.~Heck in performing CORSIKA
simulations and of N.~Kalthoff and M.~Kohler in advising the radio sounding technique is
gratefully acknowledged. One of the authors (MR) is supported by the Alexander von Humboldt Foundation. 

\section*{Appendix: Parameterizations of the Atmospheric Depth}
\label{sec.app}

\begin{table}[h]
\caption{Parameters of the US-StdA~\cite{heckcorsika}.}
\label{tab:stdatmos_param}
\begin{center}
\begin{tabular}{|c|c|c|c|c|}\hline
 \textbf{Layer $i$} & \textbf{Altitude $h$} & $a_i$ & $b_i$ & $c_i$ \\ 
  & (km) & (g/cm$^2$) & (g/cm$^2$) & (cm) \\  \hline\hline
1 & 0 ...  4  & -186.5562 & 1222.6562 & 994186.38 \\
2 & 4 ... 10  & -94.919 & 1144.9069 & 878153.55 \\
3 & 10 ... 40  & 0.61289 & 1305.5948 & 636143.04 \\
4 & 40 ... 100 & 0.0 & 540.1778 & 772170.16 \\ \hline
5 &  $>$ 100 & 0.01128292 & 1. & 10$^9$ \\ 
\hline
\end{tabular}
\end{center}
\end{table}

For CORSIKA versions 5.8 (release August 1998) and higher, it is possible to read in external atmospheric models.
This option enables not only the change of the parameters but also the variable selection of the 
boundaries for the four lowest layers.

\begin{table}[htbp]
\caption{Parameters of the average Argentine winter I atmosphere.}
\label{tab:argwin1_param}
\begin{center}
\begin{tabular}{|c|c|c|c|c|}\hline
 \textbf{Layer $i$} & \textbf{Altitude $h$} & $a_i$ & $b_i$ & $c_i$ \\ 
  & (km) & (g/cm$^2$) & (g/cm$^2$) & (cm) \\  \hline\hline
1 & 0 ...  8  & -150.247839 & 1198.5972 & 945766.30 \\
2 & 8 ... 18.1  & -6.66194377 & 1198.8796 & 681780.12 \\
3 & 18.1 ... 34.5  & 0.94880452 & 1419.4152 & 620224.52 \\
4 & 34.5 ... 100 & 4.8966557223$\cdot 10^{-4}$ & 730.6380 & 728157.92 \\ \hline
5 &  $>$ 100 & 0.01128292 & 1. & 10$^9$ \\ 
\hline
\end{tabular}
\end{center}
\end{table}

\begin{table}[htbp]
\caption{Parameters of the average Argentine winter II atmosphere.}
\label{tab:argwin2_param}
\begin{center}
\begin{tabular}{|c|c|c|c|c|}\hline
 \textbf{Layer $i$} & \textbf{Altitude $h$} & $a_i$ & $b_i$ & $c_i$ \\ 
  & (km) & (g/cm$^2$) & (g/cm$^2$) & (cm) \\  \hline\hline
1 & 0 ...  8.3  & -126.110950 & 1179.5010 & 939228.66 \\
2 & 8.3 ... 12.9  & -47.6124452 & 1172.4883 & 787969.34 \\
3 & 12.9 ... 34  & 1.00758296 & 1437.4911 & 620008.53 \\
4 & 34 ... 100 & 5.1046180899$\cdot 10^{-4}$ & 761.3281 & 724585.33 \\ \hline
5 &  $>$ 100 & 0.01128292 & 1. & 10$^9$ \\ 
\hline
\end{tabular}
\end{center}
\end{table}

\begin{table}[htbp]
\caption{Parameters of the average Argentine spring atmosphere.}
\label{tab:argsp_param}
\begin{center}
\begin{tabular}{|c|c|c|c|c|}\hline
 \textbf{Layer $i$} & \textbf{Altitude $h$} & $a_i$ & $b_i$ & $c_i$ \\ 
  & (km) & (g/cm$^2$) & (g/cm$^2$) & (cm) \\  \hline\hline
1 & 0 ...  5.9  & -159.683519 & 1202.8804 & 977139.52 \\
2 & 5.9 ... 12  & -79.5570480 & 1148.6275 & 858087.01 \\
3 & 12 ... 34.5  & 0.98914795 & 1432.0312 & 614451.60 \\
4 & 34.5 ... 100 & 4.87191289$\cdot 10^{-4}$ & 696.42788 & 730875.73 \\ \hline
5 &  $>$ 100 & 0.01128292 & 1. & 10$^9$ \\ 
\hline
\end{tabular}
\end{center}
\end{table}

\begin{table}[htbp]
\caption{Parameters of the average Argentine summer atmosphere.}
\label{tab:argsum_param}
\begin{center}
\begin{tabular}{|c|c|c|c|c|}\hline
 \textbf{Layer $i$} & \textbf{Altitude $h$} & $a_i$ & $b_i$ & $c_i$ \\ 
  & (km) & (g/cm$^2$) & (g/cm$^2$) & (cm) \\  \hline\hline
1 & 0 ...  9  & -136.562242 & 1175.3347 & 986169.72 \\
2 & 9 ... 14.6  & -44.2165390 & 1180.3694 & 793171.45 \\
3 & 14.6 ... 33  & 1.37778789 & 1614.5404 & 600120.97 \\
4 & 33 ... 100 & 5.06583365$\cdot 10^{-4}$ & 755.56438 & 725247.87 \\ \hline
5 &  $>$ 100 & 0.01128292 & 1. & 10$^9$ \\ 
\hline
\end{tabular}
\end{center}
\end{table}

\begin{table}[htbp]
\caption{Parameters of the average Argentine autumn atmosphere.}
\label{tab:argau_param}
\begin{center}
\begin{tabular}{|c|c|c|c|c|}\hline
 \textbf{Layer $i$} & \textbf{Altitude $h$} & $a_i$ & $b_i$ & $c_i$ \\ 
  & (km) & (g/cm$^2$) & (g/cm$^2$) & (cm) \\  \hline\hline
1 & 0 ...  8  & -149.305029 & 1196.9290 & 985241.10 \\
2 & 8 ... 13  & -59.771936 & 1173.2537 & 819245.00 \\
3 & 13 ... 33.5  & 1.17357181 & 1502.1837 & 611220.86 \\
4 & 33.5 ... 100 & 5.03287179$\cdot 10^{-4}$ & 750.89705 & 725797.06 \\ \hline
5 &  $>$ 100 & 0.01128292 & 1. & 10$^9$ \\ 
\hline
\end{tabular}
\end{center}
\end{table}

\begin{table}[htbp]
\caption{Parameters of the US-StdA obtained with the method applied in this work, for
details see~\cite{keilhauerFZKA}.}
\label{tab:us_param}
\begin{center}
\begin{tabular}{|c|c|c|c|c|}\hline
 \textbf{Layer $i$} & \textbf{Altitude $h$} & $a_i$ & $b_i$ & $c_i$ \\ 
  & (km) & (g/cm$^2$) & (g/cm$^2$) & (cm) \\  \hline\hline
1 & 0 ...  7  & -149.801663 & 1183.6071 & 954248.34 \\
2 & 7 ... 11.4  & -57.932486 & 1143.0425 & 800005.34 \\
3 & 11.4 ... 37  & 0.63631894 & 1322.9748 & 629568.93 \\
4 & 37 ... 100 & 4.35453690$\cdot 10^{-4}$ & 655.67307 & 737521.77 \\ \hline
5 &  $>$ 100 & 0.01128292 & 1. & 10$^9$ \\ 
\hline
\end{tabular}
\end{center}
\end{table}

\end{document}